\let\accentvec\vec
\documentclass[structabstract]{aa} 

\let\vec\accentvec
\let\epsilon\varepsilon
\let\theta\vartheta
\usepackage{graphicx}
\usepackage{rotating}
\usepackage{enumerate}
\usepackage[english]{babel}
\usepackage[T1]{fontenc}
\usepackage[dvipsnames]{xcolor}
\usepackage{url,hyperref}
\hypersetup{
    colorlinks=true,   
    linkcolor=Lavender,          
    citecolor=PineGreen,        
    filecolor=magenta,      
    urlcolor=blue           
}
\usepackage{amsmath}
\usepackage{txfonts}
\usepackage{natbib}
\bibpunct{(}{)}{;}{a}{}{,}

\newcommand{\pks}{{PKS\,1424+240}}

\newcommand{\veritas}{\textit{VERITAS}}
\newcommand{\magic}{\textit{MAGIC}}

\newcommand{\fermilat}{\textit{Fermi-LAT}}
\begin{document}
\title{Luminous and high-frequency peaked blazars: \\ the origin of the $\gamma$-ray emission from \pks}

   \author{M. Cerruti
          \inst{1}
          \and W. Benbow\inst{2} \and X. Chen\inst{3,4} \and J. P. Dumm\inst{5} \and L. F. Fortson\inst{6,7} \and K. Shahinyan\inst{6,7} 
          }

   \institute{Sorbonne Universit\'{e}s, UPMC, Universit\'{e} Paris Diderot, Sorbonne Paris Cit\'{e}, CNRS, LPNHE, 4 place Jussieu, F-75252, Paris, France, 
                \email{matteo.cerruti@lpnhe.in2p3.fr}
                \and Harvard-Smithsonian Center for Astrophysics, 60 Garden Street, Cambridge, MA 02138, USA
        \and Institute of Physics and Astronomy, University of Potsdam, D-14476 Potsdam-Golm, Germany 
        \and DESY, Platanenallee 6, D-15738 Zeuthen, Germany 
        \and Oskar Klein Centre and Dept. of Physics, Stockholm University, SE-10691 Stockholm, Sweden  
        \and School of Physics and Astronomy, University of Minnesota, Minneapolis, MN 55455, USA 
        \and Minnesota Institute for Astrophysics (MIfA), University of Minnesota, Minneapolis, MN 55455, USA
   }   

 \date{}
 
    \abstract
   {The current generation of ground-based Cherenkov telescopes, together with the LAT instrument on-board the Fermi satellite, have greatly increased our knowledge of $\gamma$-ray blazars. Among them, the high-frequency-peaked BL Lacertae object (HBL) \pks\ (z $\simeq$ 0.6) is the farthest persistent emitter of very-high-energy (VHE; E $\geq$ 100 GeV) $\gamma$-ray photons. Current emission models can satisfactorily reproduce typical blazar emission assuming that the dominant emission process is synchrotron-self-Compton (SSC) in HBLs; and external-inverse-Compton (EIC) in low-frequency-peaked BL Lacertae objects and flat-spectrum-radio-quasars. Alternatively, hadronic models are also able to correctly reproduce the $\gamma$-ray emission from blazars, although they are in general disfavored for bright quasars and rapid flares.}
   {The blazar \pks\ is a rare example of a luminous HBL, and we aim to determine which is the emission process most likely responsible for its $\gamma$-ray emission. This will impact more generally our comprehension of blazar emission models, and how they are related to the luminosity of the source and the peak frequency of the spectral energy distribution.}
   {We have investigated different blazar emission models applied to the spectral energy distribution of \pks. Among leptonic models, we study a one-zone SSC model (including a systematic study of the parameter space), a two-zone SSC model, and an EIC model. We then investigated a blazar hadronic model, and finally a scenario in which the $\gamma$-ray emission is associated with cascades in the line-of-sight produced by cosmic rays from the source.} 
   {After a systematic study of the parameter space of the one-zone SSC model, we conclude that this scenario is not compatible with $\gamma$-ray observations of \pks. A two-zone SSC scenario can alleviate this issue, as well as an EIC solution. For the latter, the external photon field is assumed to be the infra-red radiation from the dusty torus, otherwise the VHE $\gamma$-ray emission would have been significantly absorbed. Alternatively, hadronic models can satisfactorily reproduce the $\gamma$-ray emission from \pks, both as in-source emission and as cascade emission.}
  {}

   \maketitle
  
   \section{Introduction}
 
\begin{figure*}[t!]
\begin{center}
\includegraphics[width=400pt]{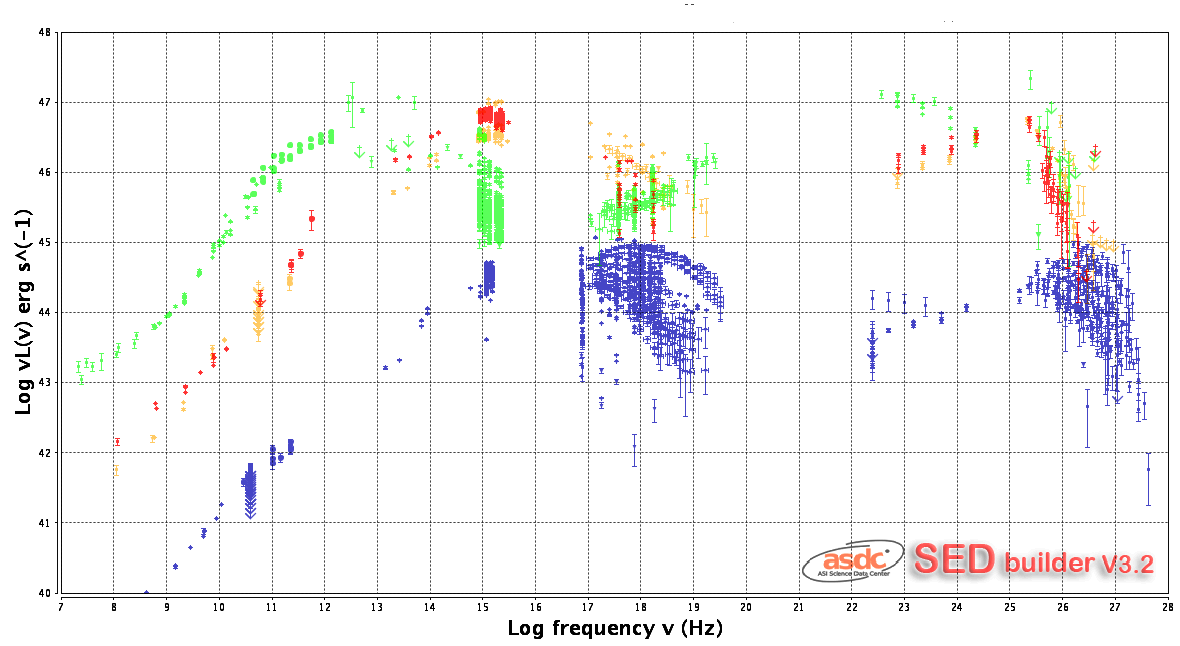}
\caption{Archival SEDs ($\nu L_\nu$ in erg/s as a function of the rest-frame frequency in Hz) of four iconic VHE blazars: 3C 279 in green, \pks\ in red, PG 1551+113 in yellow and Mrk 421 in blue. The SEDs have been compiled using the ASDC SED builder tool (\url{http://tools.asdc.asi.it/SED}). Among these four sources, 3C 279 is the only one that has been detected at VHE only during flaring activity. \label{figasdc}}
\end{center}
\end{figure*}

  Imaging atmospheric Cherenkov telescopes (IACTs) are capable of detecting astrophysical sources at energies above 100 GeV (very-high-energy $\gamma$-rays, or VHE), and have effectively opened a new window in the electromagnetic spectrum. Since the discovery of the first VHE extragalactic emitter  \citep[Markarian\ 421,][]{Punch}, the number of VHE-detected extragalactic sources has continuously increased, and nowadays 67 of them are known.\footnote{See \url{http://tevcat.uchicago.edu} for an up-to-date VHE catalog.} The great majority of VHE extragalactic sources are active galactic nuclei (AGN) of the blazar type, similar to what is observed at lower energies (0.1-100 GeV, or high-energy $\gamma$-rays, HE) by \fermilat\ \citep[see][for the most recent catalog of HE $\gamma$-ray sources]{3FGL}.\\

 A difficult aspect of extragalactic VHE astronomy is its inherent redshift limitation: the $\gamma$-ray photons emitted from the blazar can pair-produce on a low-energy (infrared and optical) photon from the diffuse extragalactic background light (EBL). This absorption effect increases with the distance and with the energy of the VHE photon: the result is a softening of blazar spectra for increasing redshifts, and, finally, a limit on the detection of distant blazars above 100 GeV \citep{Salamon98}. On the other hand, this absorption can be used to put constraints on the EBL itself and indeed, the VHE detection of the first blazars at redshifts $z \ge 0.15$ imposed strong constraints on early EBL models \citep{HESSEBL2006}. Nowadays, among the VHE blazars with a spectroscopic redshift, the most distant ones are the quasars S3~0218+35 \citep[z=0.944,][]{S30218} and PKS 1441+25 \citep[z=0.939,][]{1441MAGIC, 1441VERITAS}. Both of these quasars have been detected at VHE only during flaring activity. The most distant persistent source of VHE photons, again among the VHE blazars with a spectroscopic redshift, is \pks, observed by both \veritas\ \citep{1424discovery, 1424veritas} and \magic\ \citep{1424magic}. A first firm lower limit on its distance has been established by  \citet{Furniss13} as z $\geq$ 0.6035. \citet{Rovero16} associated this blazar with a galaxy cluster at z = 0.601 $\pm$ 0.003, in agreement with the lower limit. More recently, \citet{Paiano17} estimated the redshift of \pks\ as  z=0.604 from the detection of faint emission lines in its optical spectrum. In the following we have adopted z=0.6 as the redshift of the object, and we use the EBL model by \citet{Franceschini08}.   \\

   Blazars are radio-loud AGN characterized by a broad non-thermal continuum from radio to $\gamma$-rays, rapid variability and a high degree of polarization \citep[see e.g.,][]{Angel80}.
 These observational properties are explained as the non-thermal emission from a relativistic jet pointed in the direction of the observer \citep{Blanford78}. The blazar class is further divided into the two-subclasses of BL Lacertae objects and flat-spectrum-radio-quasars (FSRQs) according to the absence (in the former) or presence (in the latter) of emission lines in their optical spectrum \citep[see e.g.,][]{Stickel91}. These two sub-classes are also characterized by different luminosity and redshift distributions \citep[see][]{Padovani92, Massaro09} and are considered as the blazar version of the two radio-galaxy subclasses defined by \citet{Fanaroff74}.\\

 The spectral-energy-distribution (SED) of blazars is always comprised of two bumps, peaking in mm-to-X-rays and MeV-to-TeV, respectively. While FSRQs are in general characterized by the first peak at lower frequencies (in infrared), BL Lac objects have different first peak frequencies, and are thus further classified into low/intermediate/high-frequency-peaked BL Lac objects (LBLs, IBLs, HBLs, with a first peak frequency below $10^{14}$ Hz, at $10^{14-15}$ Hz, or above $10^{15}$ Hz, respectively). The majority of VHE blazars are indeed HBLs \citep[for a recent review on VHE results, see][]{denaurois15}.\\

 The different blazar subclasses (FSRQs, LBLs, IBLs, HBLs) are characterized not only by different frequencies of the synchrotron peak, but also by different luminosities. \citet{Fossati98} proposed the existence of a "blazar sequence" characterized by an anticorrelation between luminosity and peak frequency. The most powerful blazars, but with the lowest peak frequencies, would thus be the FSRQs, while the least luminous blazars would be the HBLs. Several authors  \citep[see e.g][]{Ghisellini08, Nieppola08, Giommi12} have investigated whether this sequence is real or due to selection effects. \citet{Padovani03, Caccianiga04, Padovani12} have shown that there exist sources which break the blazar sequence (i.e., powerful HBLs or low-luminosity FSRQs). In Fig. \ref{figasdc} we reproduce the average SEDs of \pks\ together with three well known VHE blazars: the FSRQ 3C~279 (z = 0.5362), the HBL PG~1553+113 (assuming z $\simeq$ 0.5), and the HBL Mrk~421 (z =  0.031).  It is clear that the peak luminosity of \pks\ is at the same level as 3C~279, but with a peak frequency two orders of magnitude higher. Compared to Mrk 421, \pks\ has a peak frequency a factor of 10 lower, but it is two orders of magnitude more luminous. The SED of \pks\ is indeed very similar to the SED of PG~1553+113, and both sources can be seen as examples of luminous and high-frequency peaked blazars.
However, it is important to remind the reader that the exact redshift of PG~1553+113 is still uncertain, and the current best estimates constrain it to be between 0.395 and 0.58 \citep{Danforth10}. The value of z=0.5 adopted here is compatible with the observational constraints, but the source may be closer, and thus less luminous than what is shown in Fig. \ref{figasdc}.\\

\begin{figure}[t!]
\begin{center}
\includegraphics[width=255pt]{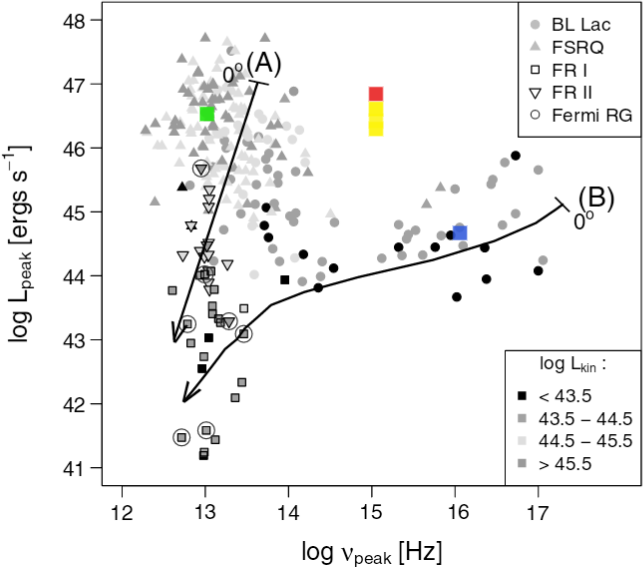}
\caption{Adapted from \citet{Meyer11}. Distribution of blazars in the $L_{peak}-\nu_{peak}$, showing the different blazar subclasses, as well as the two theoretical paths associated with a fast (A) and a slow (B) jet, observed at different angles. We overplot the position of the four VHE blazar shown in Fig. \ref{figasdc}: 3C279 in green, \pks\ in red, PG 1553+113 in yellow, and Mrk 421 in blue (for the low state). For PG 1553+113, the yellow band represents the uncertainty in its redshift. \label{figmeyer}}
\end{center}
\end{figure}

 \citet{Meyer11} have extended the blazar sequence into a "blazar envelope", in which the different blazar subclasses are due to the progressive misalignment of two intrinsically different populations of blazars (which then correspond to the Fanaroff \& Riley dichotomy in radio-galaxies). In Fig. \ref{figmeyer} we reproduce the $L_{peak}-\nu_{peak}$ plot from \citet{Meyer11}, including the values for \pks, assuming the redshift z=0.60. It is clear that \pks\ is an outlier compared to the other known blazars, and its high distance implies that it is indeed a powerful HBL, breaking the Fossati blazar sequence, and representing an intermediate blazar between the "fast-jet" and "slow-jet" blazars proposed by \citet{Meyer11}.\\
 
  Among the two components of the blazar SED, the low-energy one is clearly associated with synchrotron emission by electrons and positrons moving relativistically along the jet. The origin of the high-energy component is more disputed: in leptonic models it is associated with inverse-Compton scattering between the same $e^\pm$ population and a soft photon field, like their own synchrotron emission \citep[synchrotron-self-Compton model, SSC, see ][]{Konigl81} or an external photon field, such as the accretion disk, the broad-line-region (BLR), or the dust torus \citep[external-inverse-Compton model, EIC, see][]{Sikora94}; in hadronic models the high-energy component of the SED is instead associated with synchrotron emission by protons, and/or by secondary particles produced in $p-\gamma$ interactions \citep[see e.g.,][]{Mucke01}. Blazar emission models are not fully understood and, in general, different kinds of models are used for HBLs or LBLs/FSRQs. In the case of HBLs, the absence of emission lines and of the blue-bump associated with the accretion disk, suggests that the dominant soft photon field is the lepton synchrotron emission, and the models under study are usually limited to the SSC and the hadronic scenario. On the other hand, for LBLs and FSRQs the $\gamma$-ray emission is usually explained by EIC models \citep[see e.g.,][]{Meyer13}. In addition, while hadronic models are in general disfavored for FSRQs \citep{Sikora09, Petropoulou15, zdz15}, they can correctly reproduce the SED of HBLs (with the exception of rapid flaring activity). An intermediate object as \pks\ can thus be very useful to provide insights on the transition between the different blazar subclasses, and ultimately cast light on the physics of relativistic jets from super-massive black holes.\\
  
   Before presenting in details the modeling of \pks, it is interesting to discuss the case of PG~1553+113 which, as presented above, shows a SED similar to the one from \pks. Different authors have successfully modeled its emission using a simple one-zone SSC model \citep{Albert07, Abdo10, Aleksic10, Tavecchio10, Zhang12}. However, the uncertainty in the distance of this object complicates the modeling task and indeed most solutions adopted redshift values which have later been proven to be incorrect. \citet{Albert07} successfully modeled the SED assuming z=0.3, and found that for z>0.56 the emission from PG~1553+113 is not compatible with a simple one-zone SSC model. \citet{Aleksic10} and \citet{Zhang12} also presented a successful one-zone SSC model assuming z=0.3, while \citet{Tavecchio10} assumed z=0.36. \citet{Abdo10} successfully modeled the SED of PG~1553+113 assuming z=0.75, but they needed unusual values for the emitting region size (R$\simeq$10$^{18}$ cm, that is a factor of at least ten larger than usual SSC modeling) and an ad-hoc double-broken power-law distribution for the electrons. Another VHE blazar potentially similar to \pks\ is KUV~00311-1938 \citep{Becherini12}, but in this case as well the redshift is not well determined and only a firm lower limit of z>0.51 exists \citep{Pita14}.\\ 
 
An alternative to standard blazar emission models is represented by radiative processes in the path from the source to the observer. If the blazar is capable of accelerating ultra-high-energy cosmic rays (UHECRs), their interaction with low-energy photon fields while travelling to the Earth may be detected as an additional $\gamma$-ray component which suffers a significantly lower EBL attenuation. This UHECR origin of VHE photons has been proposed by several authors such as \citet{Essey10, Murase12}. For a specific application of this scenario to \pks\ see \citet{essey14} and  \citet{Yancascade}. \\
 
  The most recent $\gamma$-ray observations of \pks\ have been presented by \citet{1424veritas} and \citet{1424magic}, including multi-wavelength observations in optical and X-rays. In this paper we model the SED of \pks\ from \citet{1424veritas} in the framework of the standard stationary blazar leptonic and hadronic models, trying to constrain the particle content of the emitting region and its physical properties.\\

  \section{Leptonic models}  
  \subsection{Synchrotron-self-Compton model}
\label{sectionssc}  
  
   \begin{figure*}[t!]
\begin{center}
\includegraphics[width=350pt]{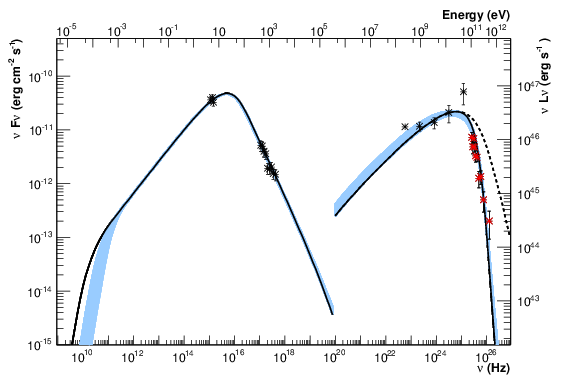}
\caption{One-zone SSC modeling of \pks. The blue band represents all of the SSC models corresponding to the parameters given in Table \ref{tablessc}. The black model represents the SSC solution provided by \citet{Kang16}. The black dashed line in the VHE regime represents the solution provided by \citet{Kang16} computed at the source, before absorption on the EBL. \label{figuressc}}
\end{center}
\end{figure*}  

 The simplest blazar emission model is the one-zone synchrotron self-Compton one \citep{Konigl81}, that correctly describes the SED of HBLs in their stationary state. The model assumes that the emission is dominated by one plasmoid propagating relativistically in the AGN jet. It is parametrized by its radius $R$ (assuming a spherical geometry), its Doppler factor $\delta$ (relative to the observer) and it is assumed to be filled by a homogeneous, tangled, magnetic field $B$. The stationary particle population energy distribution is non-thermal and parametrized by a broken-power-law function (to take into account the break associated with synchrotron losses). The particle energy distributions adds six additional parameters: the two indices $\alpha_{1,2}$, the Lorentz factors $\gamma_{min, br, max}$ and the normalization factor $K$. The number of free parameters is thus nine, and the model can be constrained only if both SED peaks are well measured. The standard approach to constrain the SSC model parameter space is the one developed by \citet{Tavecchio98}: the peak frequencies and luminosities can be analytically expressed as functions of the model parameters, and thus can be used to constrain them. The current generation of $\gamma$-ray telescopes has however greatly improved the measurement of the $\gamma$-ray SED peak, and recently, more advanced fitting and constraining algorithms have been developed by several authors \citep{Finke08, Manku11, Zhang12, CerrutiSSC}.\\
 
 In this work we make use of the constraining algorithm described in \citet{CerrutiSSC}, which is a numerical extension of the analytical work by \citet{Tavecchio98}, taking into account the GeV and TeV measurements. We simulated $10^5$ theoretical SEDs, spanning the following parameter space~: $\delta \in [30,500]$; $B \in [0.001,0.05]$ G; $R \in [10^{15}, 2\times10^{17}]$; $K \in [5\times10^{-8}, 3\times10^{-5}]$; $\gamma_{br} \in [10^4, 10^5]$. We assumed $\alpha_1 = 1.8 $, $\alpha_2 = 5.0$, $\gamma_{min}=100$, $\gamma_{max}=5\times 10^6$ (the minimum and maximum Lorentz factors of the electron distribution do not affect the SED as long as they are low and high enough). For each theoretical SED we estimated the flux and frequency of the synchrotron peak, the flux and spectral index in the \fermilat\ energy band, and the 
 flux and spectral index in the IACT energy band. For the IACT observables, we used the results obtained by  \citet{1424veritas} during the 2013 campaign, that is a detection between 100 and 750 GeV, and a decorrelation energy of 200 GeV. The next-step in the constraining algorithm would be to parametrize every observable as a function of the model parameters, and solve the system for the observable values measured for \pks. However we realized that none of the simulated SEDs has a TeV index compatible with the \veritas\ one ($4.5 \pm 0.2$): all of them are systematically softer, and range between $5.0$ and $7.5$. To investigate further the SSC modeling, we then run the constraining algorithm assuming a maximum arbitrary index in the VHE regime of $6.0$. In this case the algorithm converges and the results are given in Table \ref{tablessc}.\\

        \begin{table}
   \centering
   \caption{Parameters used for the one-zone SSC modeling of \pks }
   		\begin{tabular}{l c }
		\hline
		\hline
		 \noalign{\smallskip}
		& 2013 \\
		 \noalign{\smallskip}
		\hline
		 \noalign{\smallskip}
	z & 0.6  \\
		$\delta$ & $>250$  \\
$R_{src}$ [10$^{16}$ cm] & $0.2-1.6$ \\
 \noalign{\smallskip}
		\hline
		 \noalign{\smallskip}
		$B$ [mG] &  $1.6-8.1$  \\
		$^\star u_B\ [10^{-5}\ \textrm{erg cm}^{\textrm{-3}}\textrm{]}$ &  $0.01-0.26$ \\
		 \noalign{\smallskip}
		\hline
		 \noalign{\smallskip}
		$\gamma_{e,min}\ $&  $100$ \\
		$\gamma_{e,break}\ [10^4]$&$2.4-3.8$   \\
		$\gamma_{e,max}\ [10^6]$&  $5$ \\
		$\alpha_{e,1}$ & $1.8$\\
		$\alpha_{e,2}$ & $5.0$\\
		$K_e\ [10^{2}\ \textrm{cm}^{\textrm{-3}}\textrm{]}$  & $1.4-22.1$\\
		$^\star u_e\ [10^{-3}\ \textrm{erg cm}^{\textrm{-3}}\textrm{]}$  & $3.6-51.0$\\
		 \noalign{\smallskip}
		\hline
		 \noalign{\smallskip}
   	$^\star u_e/u_B\ [10^3] $ & $1.7-130$  \\
		$^\star L$ [10$^{45}$ erg s$^{\textrm{-1}}$ ]&  $2.5-11.8$  \\
		 \noalign{\smallskip}
		\hline
		\hline		
		 \noalign{\bigskip}	
		\end{tabular}
	 \label{tablessc}
	 \newline
		 The luminosity of the emitting region has been calculated as $L=2 \pi R^2c\Gamma_{bulk}^2(u_B+u_e+u_p)$, where $ \Gamma_{bulk}=\delta/2$. The energy densities of the magnetic field, the electrons, and the protons, are indicated as $u_B$, $u_e$, and $u_p$, respectively.  The quantities flagged with a star are derived quantities and not model parameters.
		\end{table}

 For every SSC solution we recompute the corresponding model which is shown in Fig. \ref{figuressc}. Looking at the parameters, it is clear that the critical one is the Doppler factor. A minimum value of $\delta = 250$ is extremely high (by one order of magnitude) compared to the solutions achieved for other blazars \citep{Tavecchio10, Zhang12}. This result, together with the fact that these solutions are all systematically softer than the VHE measurement, implies that the simple one-zone SSC model is highly disfavored to explain the $\gamma$-ray emission from \pks.\\
      
 Only two other SSC modeling attempts of \pks\ are available in the literature, after its redshift of $0.6$ has been determined. \citet{Kang16} has modeled the same 2013 \veritas\ campaign using a $\chi^2$ minimization algorithm, finding a solution with $\delta=51$ and $B=0.02$, which is not included in our set of solutions. To understand this issue, and compare the results from different fitting algorithms, in Fig. \ref{figuressc} we plot their solution as well, reproduced with our numerical code using their model parameters. By fitting its corresponding spectral index in the VHE regime, we find that the solution is excluded by our algorithm because it has a $\Gamma_{VHE} = 6.4$, slightly softer than the limit we adopted. \citet{Kang16} also disfavor the SSC model, because it has a reduced $\chi^2$ value of 2.2. For comparison, our best reduced $\chi^2$ is 1.6, but again, achieved only for extreme values of $\delta$. Probably, our solutions with $\delta>250$ are not reported by \citet{Kang16} due to a reduction of their parameter space to reasonable parameter values.\\   
		
\citet{1424magic} also presented a one-zone SSC modeling of \pks\ after the new constraint on its redshift: they described the blazar SED assuming an extreme Doppler factor of $70$. However, it should be noted that the \magic\ spectrum has a lower exposure compared to the \veritas\ one, and that their last significant bin is at an energy of 400 GeV only. We thus re-run our constraining algorithm considering a smaller energy-range (150-400 GeV): the distribution of simulated VHE indices, as expected, becomes harder, and there exist indeed models compatible with the \magic\ spectral index ($\Gamma_{VHE} = 5.0 \pm 1.7$). Anyhow, the modeling by \citet{1424magic} also disfavors an SSC origin of the $\gamma$-ray photons from \pks, due to the high value of $\delta$. The \veritas\ detection up to 750 GeV is even more constraining: there are no SSC models which can reproduce the emission from \pks, and even when allowing models with $\Gamma_{VHE} = 6.0$, only solutions with unrealistic Doppler factors are found.\\ 

A possible solution to this problem is that the dominant emission process is SSC, but that the acceleration and radiation mechanisms are more complex with, for example, several emitting regions contributing to the $\gamma$-ray component. We investigated this option by using the numerical code described in \citet{Chen15}, which simulates the emission from particles both in an acceleration zone and a diffusion zone. The diffusion zone is a much larger spherical zone surrounding the acceleration zone. Particles are only injected and accelerated in the acceleration zone, but they are subject to spatial diffusion and radiative cooling in both zones. The mechanism of acceleration and injection of the particles are conjectured to be magnetic reconnection or second-order Fermi acceleration, whereas the particles in the diffusion zone are the particles escaped from the acceleration zone through spatial diffusion.\\
 
\begin{figure}[t!]
\begin{center}
\includegraphics[width=260pt]{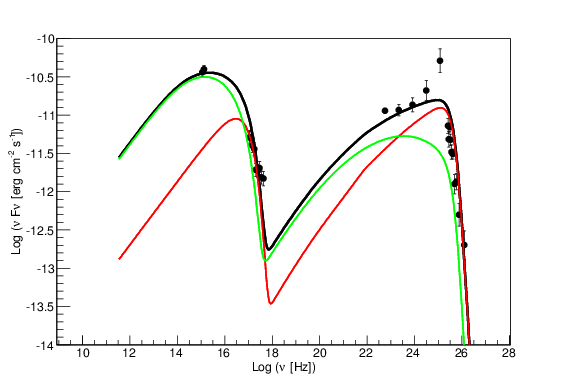}
\caption{Two-zone SSC modeling of \pks, using the model described in \citet{Chen15}. The green line represents the emission from the accelerating region, and the red one the emission from the diffusion region. The black line represents the combined emission from both zones.\label{figurexuhui}}
\end{center}
\end{figure}

In Fig. \ref{figurexuhui} we present the modeling of the \pks\ SED during the 2013 campaign in a two-zone scenario. In this case, a good description of the $\gamma$-ray emission can be achieved assuming $\delta=35$, $B=0.033$ G, $R_{acc}=5\times10^{16}$ cm and $R_{diff}=4\times10^{17}$ cm, where $R_{acc}$ and $R_{diff}$ are the radii of the acceleration and diffusion regions, respectively. Contrary to the one-zone SSC modeling, the information from the VHE spectral index is not explicitly used in the two-zone SSC modeling. To ease the comparison between the results achieved with the one-zone and the multi-zone models, we thus performed a fit of the two-zone SSC model within the \veritas\ detection range. The result is a spectral index $\Gamma_{VHE}=5.2$. This index is much closer to the measured \veritas\ index ($4.5 \pm 0.2$) than in the one-zone SSC scenario. In addition, the value of $\delta=35$ is much lower than the values obtained in the one-zone SSC modeling, and in line with typical blazar values.\\

\begin{table*}[hbtp]
\begin{center}
\caption{Parameters used for the EIC modeling of \pks}
\label{tableone}
\begin{tabular}{c|c c c c c c c| c c c c c c c }
\hline
\hline
 & \multicolumn{7}{c|}{Input} & \multicolumn{7}{c}{Output} \\
\hline
Epoch$^a$ & $L_{48}$ & $t_4$ & $\nu_{14}$ & $\zeta_e$ & $\zeta_s$ & $\zeta_{ext}$ & $b$ & $\delta$ & $B$ & $R$ & $\gamma'_p$ & $N'_e(\gamma'_p)$ & $u_{ext}$ & $L_{jet}^b$ \\
 &  & &  &  & & & & & G & $10^{17}\ \textrm{cm}$ & $10^3$ & $10^{-5}\ \textrm{cm}^{-3}$ & $10^{-7}\ \textrm{erg cm}^{-3}$ & $10^{45}\ \textrm{erg s}^{-1}$\\
\hline
2009 & 0.55 & 10 & 120 & 1 & 0.8 & 0.5  & 0.6 & 30 & 0.14 & 0.9 & 3.8 & 1.9 & 7.8& 1.0\\
2013 & 0.38 & 10 & 25   & 1 & 0.2 & 0.5  & 0.9 & 46 & 0.07 & 1.4 & 3.9 & 0.5 & 1.3& 1.2\\
\hline
\hline
\end{tabular}
\endcenter
$^a$Data and model SEDs are shown in Fig. \ref{EICfig}.\\
$^b$Total jet luminosity assuming the energy density of hadrons equals that of electrons.
\end{center}
\end{table*}

   \begin{figure*}[hbtp]
\begin{center}
\includegraphics[width=250pt]{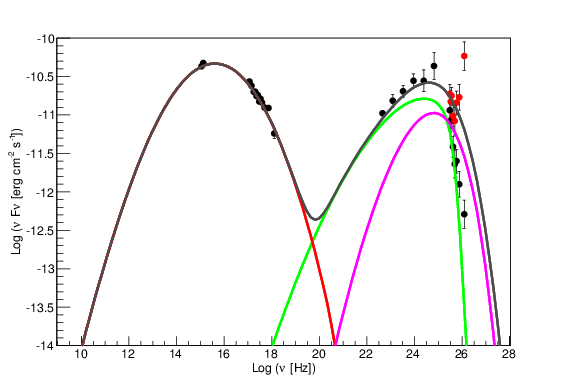}
\includegraphics[width=250pt]{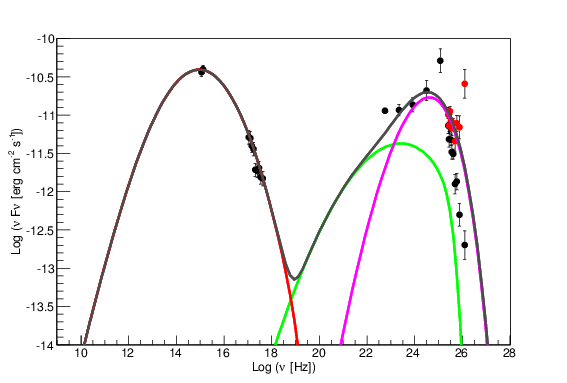}
\caption{Modeling of \pks\ in an EIC scenario, for the 2009 (\textit{left}) and 2013 (\textit{right}) campaigns. The black solid line represents the total emission; the red line represents the synchrotron emission by leptons; the green line the SSC component; the pink line the EIC component assuming that the external photon field is produced by the dusty torus. The model parameters are provided in Table \ref{tableone}. The models do not include absorption over the EBL. The \veritas\ data \citep{1424veritas} deabsorbed using the EBL model by \citet{Franceschini08} are plotted in red. \label{EICfig}}
\end{center}
\end{figure*}
 
 \subsection{External-inverse-Compton model}  
 The second scenario we investigated is an EIC model in which the leptons in the emitting region also scatter external photons from the BLR, the dust torus, and/or the accretion disk. The main reason for testing this scenario is that the EIC model correctly describes the SED of powerful FSRQs, and we expect that, if \pks\ represents indeed an intermediate blazar, it can have an external photon field bright enough to contribute to the overall photon emission. The EIC model has many more free parameters than the SSC one, and a detailed study of the parameter space cannot be performed without the addition of explicit assumptions in order to reduce the number of degrees of freedom. In particular, the energy densities of the external photon fields are free parameters that depend on the exact location of the $\gamma$-ray emitting region in the jet. To alleviate this problem, and reduce the number of free parameters, we adopted the approach described in \citet{chuckii} and \citet{chucki}: under the assumption that the emitting region is close to equipartition between magnetic field, particle, synchrotron photon and external photon energy densities ($u_B$, $u_e$, $u_s$, and $u_{ext}$, respectively), and that the particle population can be parametrized by a log-parabolic function, we determine the values for the Doppler factor $\delta$, the magnetic field $B$, the emitting region size $R$, the peak Lorentz factor of the particle energy distribution $\gamma_{p}$, and the energy density of the external photon field that correctly model the SED. The equipartition factors are defined as $\zeta_e = u_e / u_B$; $\zeta_s = u_s / u_B$; and $\zeta_{ext} = u_{ext} / u_B$.\\
 
 An important aspect of EIC scenarios is the absorption of $\gamma$-rays via pair-production on the external photon field (similarly to the absorption on the EBL discussed above). In particular, Ly$\alpha$ photons (with E=10.2 eV) from the BLR can efficiently absorb VHE photons \citep{Ghisellini09}. The very detection of VHE FSRQs implies that the $\gamma$-ray emission is located outside the BLR during the VHE detection, or at the most at its very edge, in order to escape the low-energy photon field and ultimately reach the observer \citep{3C279, 1222, 1510, 1510b, 1441VERITAS, 1441MAGIC}. For \pks\ the scenario is similar to VHE FSRQs: the detection of $\gamma$-ray photons up to 750 GeV means that the $\gamma$-ray emitting region is located outside the BLR, or that the BLR itself is extremely underluminous. In the modeling we thus consider as the external photon field only the thermal emission from the dusty torus surrounding the super-massive black hole. \\
 
  We modeled the two SED compiled by the \veritas\ collaboration in 2009 and 2013. We assume a peak frequency of the synchrotron component $\nu_{peak}=12\ (4) \times10^{15}$ Hz, and a luminosity of $\nu_{peak} L_{peak, synch}=5.5\ (3.8) \times10^{47}$ erg s$^{-1}$ for the 2009 (2013) season (see Table \ref{tableone}). The \textit{Swift-XRT} observations constrain the shape of the particle population, and we considered a curvature index $b=0.6$ and $0.9$ for the 2009 and 2013 seasons, respectively. We assumed a variability time-scale of $10^5$ seconds (27.8 hours), which matches the doubling time-scale of about one-day observed in soft X-rays during the 2009 multi-wavelength campaign \citep{1424magic}. We impose perfect equipartition between the magnetic field and the lepton energy density ($\zeta_e=1$). The values of the density of synchrotron photons and external photons are adjusted to reproduce the $\gamma$-ray emission. The SED high-energy bump is associated with SSC scattering and EIC scattering over the dusty torus thermal emission. The output parameters are reported in Table \ref{tableone}: in our scenario the 2013 activity is associated with a higher $\delta$, a lower $B$ and a larger but less dense emitting region; the jet power increases by 20\%. The external photon energy density is reduced by a factor of four, but we explain this variability not in terms of variability of the dust torus, but rather by a change in the location of the emitting region. If the emission is produced at a distance $r$<$r_{IR}$ (where $r_{IR}$ is the radius of the torus), the photon energy density should be several order of magnitudes higher than the $10^{-7}$ erg cm$^{-3}$ inferred from our modeling \citep{Ghisellini09}. This means that in our scenario the emission is produced at $r$>$r_{IR}$, where the energy density of the infrared photon field is rapidly decaying. Assuming $r_{IR} \simeq 2.5 \times 10^{18} L_{d,45}^{1/2}$ cm (where $L_{d,45}$ is the luminosity of the accretion disk, in units of $10^{45}$ erg s$^{-1}$) we can thus infer the location of the emitting region at the parsec scale from the super-massive black-hole. This study has been performed using as observables the energies and luminosities of the SED peaks, and by imposing equipartition criteria, but not using as a constraint the index in the VHE range. Following what we did for the two-zone SSC model, in this case as well we performed a fit of the EIC models  in the VHE regime. The resulting spectral indices are $\Gamma_{VHE}=6.1$ and $\Gamma_{VHE}=6.3$ for the 2009 and 2013 campaigns, respectively. These indices are softer than the observed one, and similar to the results obtained with the SSC code. However, the value of $\delta=30-46$ is significantly lower than in the one-zone SSC scenario.\\
  
\citet{Kang16} also attempted an EIC modeling of \pks. They presented a solution in which the VHE emission is dominated by EIC scattering over infrared photons from the dust torus, similar to ours. The model parameters are also similar, even though they adopted a minimization algorithm, and did not explicitly force equipartition between the energy densities of the leptons and the magnetic field energy.\\

  \section{Hadronic model}

   \begin{figure*}[hbtp]
\begin{center}
\includegraphics[width=260pt]{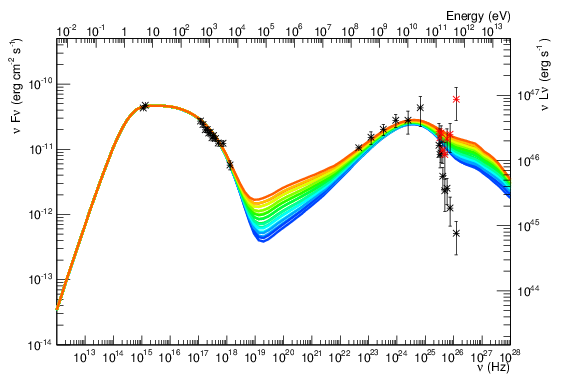}
\includegraphics[width=260pt]{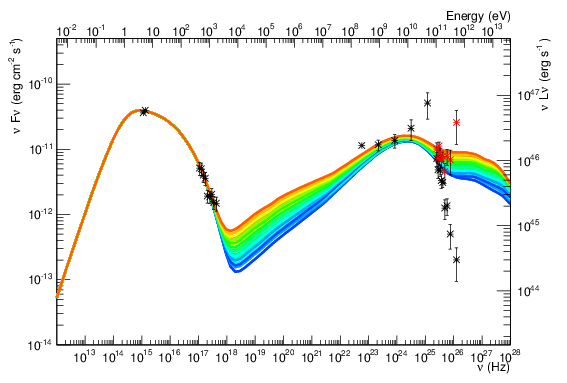}
\caption{Modeling of \pks\ in a hadronic scenario, for the 2009 (\textit{left}) and 2013 (\textit{right}) campaigns, for a value of the Doppler factor $\delta=30$. Only the overall emission is plotted. The colors, from blue to red, represent a decrease in the size of the emitting region $R$, and an increase in the strength of the magnetic field $B$.  The model parameters are provided in Table \ref{table1a}.  The \veritas\ data \citep{1424veritas} deabsorbed using the EBL model by \citet{Franceschini08} are plotted in red.\label{figurehadro}}
\end{center}
\end{figure*}

\subsection{Emission at the source}
\label{hadronicmodelsection}
  In an alternative scenario, the high-energy component of the SED is associated with protons which either directly radiate synchrotron photons, or produce secondary mesons and leptons via proton-photon interactions. The hadronic modeling is performed using the code described in \citet{leha}: p-$\gamma$ interactions are computed using the Monte-Carlo code SOPHIA \citep{sophia}, and the code correctly computes synchrotron emission by $\mu^\pm$ (before decaying back into e$^\pm$) and by synchrotron-supported pair-cascades triggered by the secondary e$^\pm$ and the photons from $\pi^0$ decay. The Bethe-Heitler pair-production (not included in SOPHIA) is calculated analytically following \citet{Kelner08}. The code does not take into account external photon fields; it assumes that the dominant low-energy photons are the synchrotron ones, and not the ones from the BLR, the torus or the accretion disk. It thus has to be compared directly to the leptonic SSC scenario, and not the EIC.\\
  
  Blazar hadronic modeling has a much higher number of free parameters, compared to the standard SSC model, due to the additional parameters associated with the proton energy distribution. The particle energy distributions for electrons (subscript $e$) and protons (subscript $p$) are parametrized as broken power-laws with indices $\alpha_{1,2}$. The relevant energies (minimum, break, and maximum) are provided as Lorentz factors $\gamma_{min, break, max}$. The normalizations of the particle energy distributions ($K_e$, and $\eta=K_p/K_e$) are provided at $\gamma=1$. Following \citet{leha}, we can reduce the number of parameters by making two assumptions on the acceleration mechanism: the first one is that the electrons and protons are co-accelerated, and they thus share the same injection index ($\alpha_{e;1} = \alpha_{e;2}$); the second one is that the maximum proton energy is constrained by the equality of the acceleration and cooling time-scales. However, even under these hypotheses, the number of free parameters is higher than the number of independent observables: we thus search for solutions only for a given set of Doppler factor values $\delta=15,\ 30,\ 60$. We first search for pure proton-synchrotron solutions, but we fail in modeling the SED. The reason is that the emission from secondary particles from p-$\gamma$ interactions is important, and significantly modify the $\gamma$-ray component. On the other hand we find good hadronic solutions dominated by proton-synchrotron emission in the MeV-GeV part of the spectrum, and by synchrotron emission from secondary leptons in the VHE regime. This type of hadronic solution is very similar to the one obtained for the HBLs Mrk 421 and PKS 2155-304 by \citet{lehacta}. The injection indices for electrons and protons are fixed to $1.8$. The parameter space is studied fixing the peak frequency of the proton-synchrotron component, and moving along a line in the $R$-$B$ plane \citep[see][ for details]{leha}.  When varying $\delta$, the parameter which compensates for the varying photon emission is the particle density: for the same values of $R$ and $B$, solutions with lower $\delta$ have thus a higher value of particle density, which means a higher contribution from secondary particles produced in p-$\gamma$ interactions. To better study the effect of the Doppler factor on the model parameters, we relaxed the constraint on the variability timescale: for $\delta=15$, no solutions are found for $\tau_{var}<6.5$ days for the 2013 campaign, the reason being that emission from secondary pairs in a small and dense emitting region overshoot the very soft X-ray emission observed with Swift/XRT; for $\delta=30, 60$, we imposed $\tau_{var} <3$ days. We found good solutions for both the 2009 and 2013 SEDs assuming $R$= $(3.2-32.4)\times\ 10^{16}$\ cm, and $B$ = $0.4-2.3$ G. For lower (respectively, higher) values of $R$, the emission in the VHE regime becomes too hard (respectively, soft) compared to the observations. The differences in the emission between the 2009 and the 2013 campaign are explained by a lower maximum particle energy (for both electrons and protons) in 2013. The value of $\delta$ affects the equipartition of the emitting region: for $\delta=30$ the emitting region is remarkably close to equipartition ($u_p / u_B \simeq 1$), while the emitting region is particle dominated for $\delta=15$, and magnetic-field dominated for $\delta=60$.  The required luminosity is $(4-19)\times\ 10^{46}$ erg s$^{-1}$, of the order of the Eddington luminosity of the super-massive black hole powering the blazar (which is $10^{47}$ erg s$^{-1}$ for a super-massive black hole mass of $10^9 M_\odot$). Given that the observed variability time-scale in X-rays during the 2009 multi-wavelength campaign was about one day, we disfavor the solutions for $\delta=15$. Solutions for $\delta=30$ are preferred over the ones for $\delta=60$ for both a lower luminosity and an equiparition factor close to unity. The hadronic modeling of \pks\ for the 2009 and 2013 campaigns is presented in Fig. \ref{figurehadro} (for $\delta=30$), and the detailed values assumed in the modeling are listed in Table \ref{table1a}.\\
  
 A hadronic modeling of \pks\ has been also presented by \citet{Yanhadro}. The authors found a model similar to ours, in which the $\gamma$-ray component is dominated by proton-synchrotron emission and synchrotron emission by secondary leptons from p-$\gamma$ interactions. On the other hand, the parameter values differ: \citet{Yanhadro} presented a solution with the same value of $\delta=30$, but a higher $B=15$ G and a smaller $R$= $5\times\ 10^{15}$\ cm. Using these parameters, we cannot fit the SED of \pks. The reason is that in their numerical simulation the synchrotron emission by muons is not computed, while this component is important in the VHE regime, and increases the hardening of the $\gamma$-ray emission \citep{lehacta}.\\ 
  
        \begin{table*}
   \centering
   \caption{Parameters used for the hadronic modeling of \pks }
   
\noindent\makebox[\textwidth]{   
   		\begin{tabular}{l| c c c | c c c}
		\hline
		\hline
		 \noalign{\smallskip}
		& \multicolumn{3}{c|}{2009} & \multicolumn{3}{c}{2013} \\
		 \noalign{\smallskip}
		\hline
		 \noalign{\smallskip}
	z & \multicolumn{3}{|c}{0.6} & \multicolumn{3}{|c}{0.6} \\
		$\delta$ & 15 & $30$ & 60 & 15 & $30$ & 60  \\
$R_{src}$ [10$^{16}$ cm] & $5.0-24.3$ & $5.0-16.2$& $3.2-32.4$ & $15.8-24.3$ & $7.9-16.2$ & $3.2-32.4$\\
$\tau_{var}$ [days] & $2.0-10.0$& $1.0-3.3$ &  $0.3-3.3$ & $6.5-10.0$ & $1.6-3.3$ & $0.3-3.3$ \\
 \noalign{\smallskip}
		\hline
		 \noalign{\smallskip}
		$B$ [G] & $0.7-2.0$ &$0.8-2.0$& $0.5-2.3$ & $0.6-0.8$ &  $0.7-1.3$ &  $0.4-2.2$ \\
		$^\star u_B\ [\textrm{erg cm}^{\textrm{-3}}\textrm{]}$ & $0.02-0.16$ & $0.03-0.16$& $0.008-0.22$ & $0.01-0.02$ &  $0.02-0.06$ & $0.007-0.19$  \\
		 \noalign{\smallskip}
		\hline
		 \noalign{\smallskip}
		$\gamma_{e,min}\ [10^3]$& $1.4-2.4$ & $1.2-2.0$ & $0.9-1.9$ & $3.3$-$3.9$ &  $1.3-1.8$ & $0.6-1.4$  \\
		$\gamma_{e,break}\ [10^3]$& \multicolumn{3}{c|}{$=\gamma_{e,min}$} & & $=\gamma_{e,min}$ &    \\
		$\gamma_{e,max}\ [10^4]$& $3.6-6.1$ & $3.0-4.6$& $2.1-4.8$  & $1.8-2.1$ & $1.2-1.5$  & $0.7-1.5$ \\
		$\alpha_{e,1}=\alpha_{p,1}$ & \multicolumn{3}{c|}{$1.8$}  & \multicolumn{3}{c}{$1.8$} \\
		$\alpha_{e,2}=\alpha_{p,2}$ & \multicolumn{3}{c|}{$2.8$}  & \multicolumn{3}{c}{$2.8$} \\
		$K_e\ \textrm{[cm}^{\textrm{-3}}\textrm{]}$ & $4.4-115.9$ &  $0.8-10.3$ & $0.02-2.9$  & $4.5-10.8$ & $1.4-7.0$ & $0.04-5.0$ \\
		$^\star u_e\ [10^{-5}\ \textrm{erg cm}^{\textrm{-3}}\textrm{]}$ & $0.9-21.1$  & $0.2-1.8$ & $0.004-0.5$  & $0.8-1.6$ & $0.2-1.1$ & $0.005-0.7$\\
		 \noalign{\smallskip}
		\hline
		 \noalign{\smallskip}
		$\gamma_{p,min}$& \multicolumn{3}{c|}{1} & \multicolumn{3}{c}{1}  \\
		$\gamma_{p,break} [10^9]$&  \multicolumn{3}{c|}{$=\gamma_{p,max}$}  & \multicolumn{3}{c}{$=\gamma_{p,max}$}  \\
		$\gamma_{p,max} [10^9]$& $6.5-10.9$ & $6.0-9.6$& $4.4-10.3$ & $8.1-9.4$ & $6.0-8.2$& $4.1-9.6$ \\
		$\eta$ & $0.025-0.039$ & $0.050-0.054$& $0.067-0.072$  & $0.053-0.060$ &$0.021-0.026$ & $0.028-0.033$ \\
		$^\star u_p\ \textrm{[erg cm}^{\textrm{-3}}\textrm{]}$ &$0.12-1.82$ &  $0.03-0.35$& $0.001-0.12$ & $0.19-0.38$ & $0.02-0.11$ & $0.0007-0.10$  \\
		 \noalign{\smallskip}
		\hline
		 \noalign{\smallskip}
   	$^\star u_p/u_B $ & $6.4-11.4$  & $0.9-2.6$ & $0.1-0.6$  & $12.9-15.1$ & $0.8-2.1$ & $0.08-0.6$ \\
		$^\star L$ [10$^{46}$ erg s$^{\textrm{-1}}$ ]& $5.3-9.0$ & $4.9-6.9$& $5.2-19.1$ & $10.9-12.5$ &  $4.0-5.4$ & $4.4-16.2$  \\
		 \noalign{\smallskip}
		\hline
		\hline		
		 \noalign{\bigskip}	
		\end{tabular}
		}
	 \label{table1a}
	 \newline
		 The luminosity of the emitting region has been calculated as $L=2 \pi R^2c\Gamma_{bulk}^2(u_B+u_e+u_p)$, where $ \Gamma_{bulk}=\delta/2$. The energy densities of the magnetic field, the electrons, and the protons, are indicated as $u_B$, $u_e$, and $u_p$, respectively.  The quantities flagged with a star are derived quantities and not model parameters.
		\end{table*}

\subsection{Cascades in the line of sight}

In order to model the potential contribution from line-of-sight cascades from UHECR interactions, we use the CRPropa3 software\footnote{CRPropa version 3: https://github.com/CRPropa/CRPropa3}. This publicly available tool includes the ability to simulate the propagation of UHECRs and calculate their interactions along the way, including pair production, photo-pion production, photodisintegration, and nuclear decay. In addition, it can track the production and propagation of the byproducts of these interactions (secondary photons, neutrinos, and electron-positron pairs) and output spectra for the primary and the secondary particles \citep[for full details see][]{batista16}. For consistency with the modeling in this paper, a small modification is made to the CRPropa3 code to include the \citet{Franceschini08} EBL model in the photon propagation chain.\\

Using CRPropa3, cosmic rays are propagated from the PKS 1424+240 distance of z = 0.6.  A pure-proton composition is assumed, as in the hadronic model described in section \ref{hadronicmodelsection}. For the secondary $\gamma$-ray emission to be viable, the strength of the intergalactic magnetic fields (B$_{IGMF}$) cannot exceed $\sim$1.4$\times$10$^{-14}$G; otherwise, deflections of the primary cosmic rays away from the line of sight are large enough for the emission from the secondary $\gamma$-rays to extend beyond the angular resolution of IACTs and become lost in the background. Following \citet{essey14,Yancascade}, we adopt magnetic fields with a strength B = 10$^{-15}$G and a correlation length of 1~Mpc. To be consistent with the treatment of the primary emission, the EBL model from \citet{Franceschini08} is used. The values of parameters used for both scenarios are listed in Table~\ref{tab:crpropa}.\\

\begin{table*}
\begin{center}
\caption{Parameters used in modeling the $\gamma$-ray data with UHECR-induced cascade emission.\label{tab:crpropa}}
\begin{tabular}{cccc}
\hline
& Hadronic 2009 & Hadronic 2013 & SSC 2013 \\
\hline
z & 0.6 & 0.6 & 0.6\\
B$_{IGMF}$ [10$^{-15}$ G] & 1 & 1 & 1\\
\hline
Composition & protons only & protons only & protons only \\
$\alpha_{p}$ & -1.8 & -1.8 & -1.8\\
E$_{p,min}$ [10$^{18}$ eV] &  0.07 & 0.07 & 0.07 \\
E$_{max}$ [10$^{18}$ eV] & 6.1 -- 8.6 & 5.2 -- 7.1 & 8 \\
$\theta_{p}$ [$^{\circ}$] & 3.8 & 3.8 & 3.8 \\
L$_{p,esc}$ [10$^{46}$ erg s$^{-1}$] & 0.55 -- 0.59 & 0.46 -- 0.55 &  1.5 -- 2.4\\
\hline
\hline
\end{tabular}
\end{center}
\end{table*}

\begin{figure*}[]
\includegraphics[width=260pt]{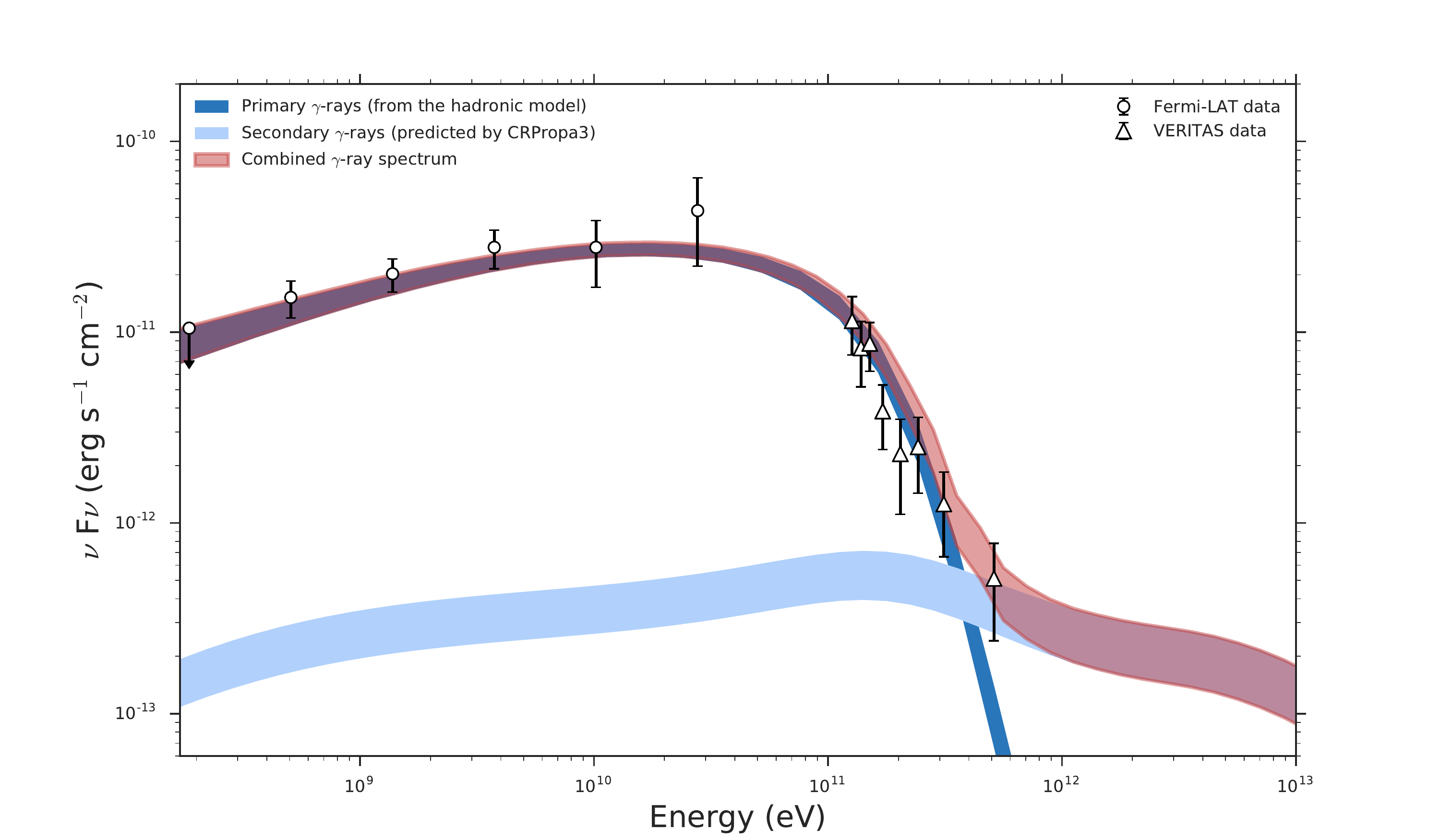}
\includegraphics[width=260pt]{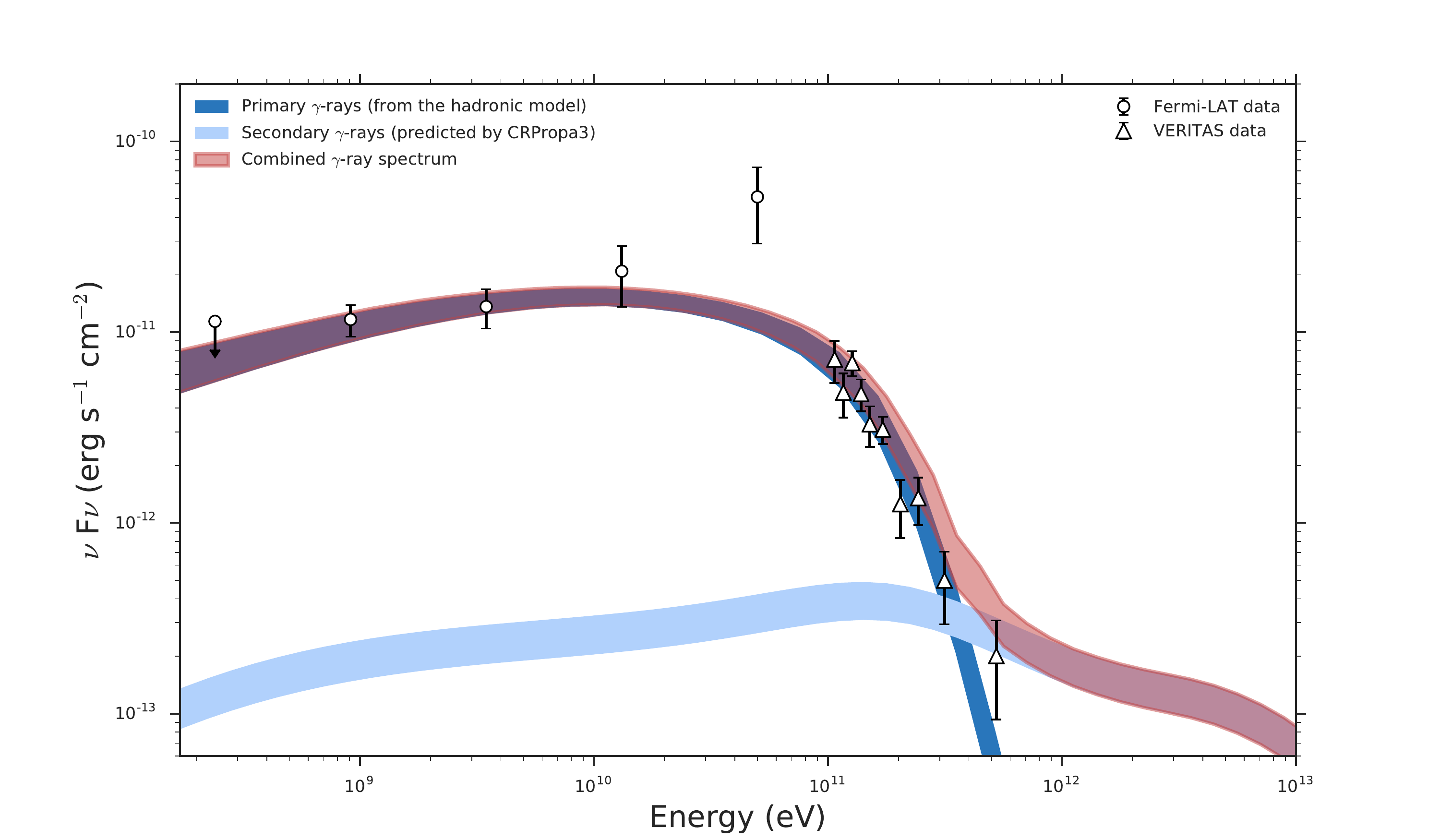}
\caption{Gamma-ray emission described by the hadronic scenario together with the predicted secondary $\gamma$-ray spectra from UHECR-initiated cascades along the line of sight for the 2009 (\textit{left}) and 2013 (\textit{right}) campaigns.}
\label{fig:crpropa_hadronic}
\end{figure*}

The spectrum of the secondary $\gamma$-rays is calculated for two scenarios with line-of-sight UHECR cascade emission as (1) an extension of the hadronic model and (2) as an additional component to the SSC model. Under the first scenario, the aim is to produce a self-consistent model of the primary and the secondary emission from PKS 1424+240. The UHECR-borne secondary $\gamma$-rays are considered within the context of the hadronic model presented in Section~\ref{hadronicmodelsection}, with parameters for the secondary emission derived from the primary model. Following the discussion in the previous section, we favor the hadronic solutions for $\delta=30$, and use the respective model parameters in the following.  Specifically, the bulk Lorentz factor ($\Gamma$), the spectral properties ($\gamma_{min}$, $\gamma_{max}$, $\alpha_{p}$), and the total power of the protons (L$_{p}$) are taken from the hadronic model and are used to constrain the range of possible normalizations of the secondary $\gamma$-ray spectra. L$_{p}$ is calculated using the total luminosity of the emitting region (L) and the equipartition parameter (u$_{p}$/u$_{B}$). It represents the isotropic power from a stationary spectrum of hadrons in the emitting region and accounts for energy losses at the source, including energy required for the production of the primary $\gamma$-rays. While during the simulation, the UHECR spectrum is cut off at 0.07~EeV at the lower energy end (due to memory concerns), below which the contribution to the secondary $\gamma$-ray emission is negligible, the full spectrum of the UHECR extending down to 1~GeV is used for calculating the power of the UHECRs required for the production of the secondary $\gamma$-rays. The Lorentz factor, $\Gamma$ = 15 determines the opening angle of the UHECRs following $\theta_{p}$ = 1/$\Gamma$ = 3.8$^{\circ}$, which translates to a factor of 900 amplification of the secondary $\gamma$-ray emission over the case where UHECR are emitted isotropically.\\

We first studied the scenario in which the entire L$_p$ used in Section~\ref{hadronicmodelsection} goes into the line-of-sight cascade. In this case the emission from UHECR cascades significantly overshoots \veritas\ observations, and we can conclude that such a self-consistent model is excluded. A first alternative is that only a fraction of the proton power goes into UHECR cascades, and we thus define an escape fraction $\xi$=L$_{p,esc}$/L$_p$. VHE observations can thus be used to put a constraint on L$_{p,esc}$, under the assumption that the broadband SED is associated with hadronic emission. It is important to underline that this result depends on the choice of B$_{IGMF}=10^{-15}$ G, and higher values of BIGMF would appreciably scatter the UHECRs and lower the observed cascade emission.\\

Fig.~\ref{fig:crpropa_hadronic} presents the full range of possible secondary $\gamma$-ray spectra calculated for $\Gamma$ and $\gamma_{max}$ values from the hadronic model applied to data from 2009 (left) and 2013 (right) campaigns. For both observing campaigns, we found that a self-consistent hadronic model can be achieved assuming a proton escape fraction of one third.\\

\begin{figure}[]
\includegraphics[width=260pt]{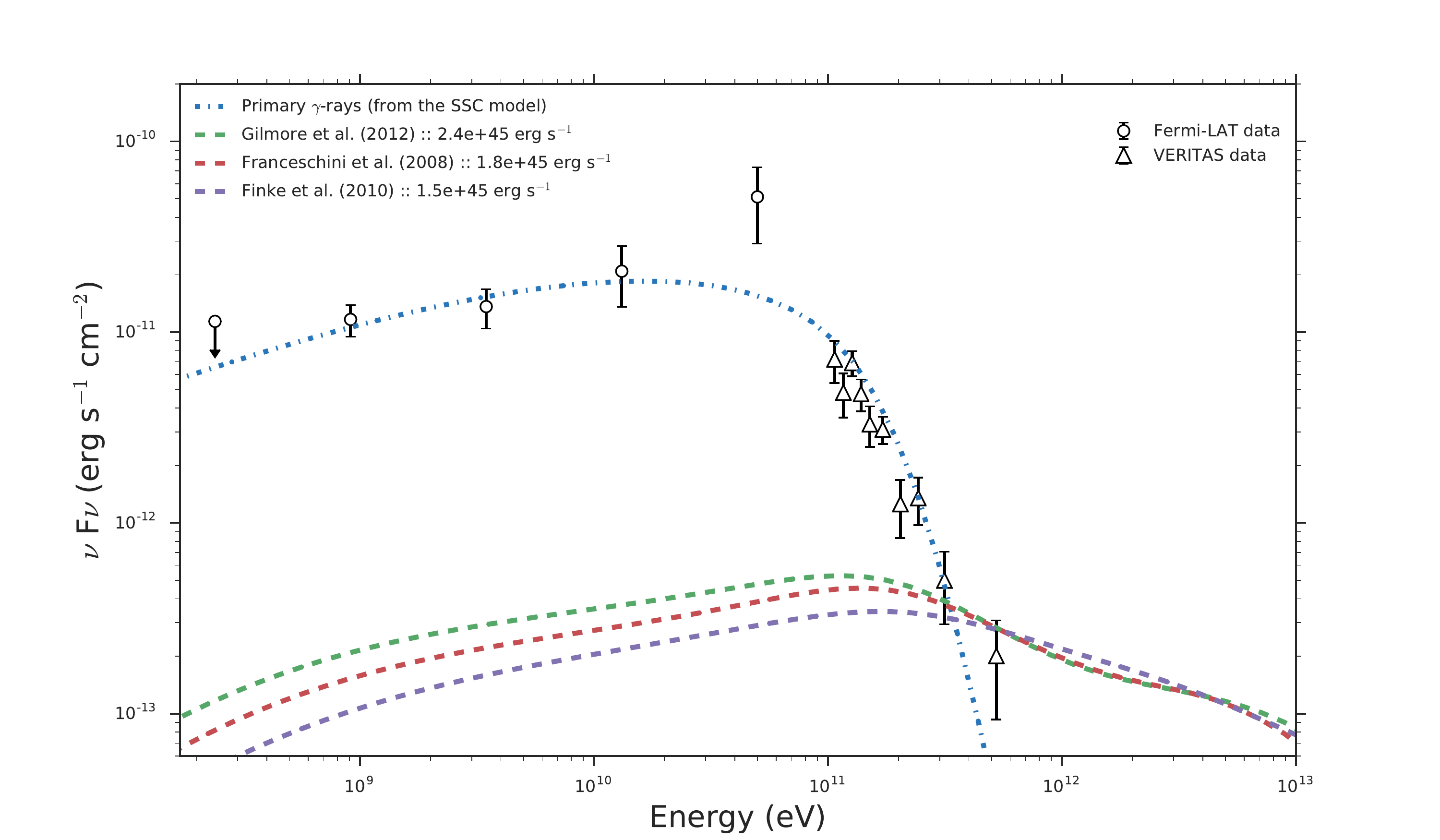}
\caption{Predictions for secondary $\gamma$-ray spectra using different EBL models within the SSC scenario for the 2013 campaign. The required UHECR power for producing the secondary $\gamma$-rays is included in the legend for each EBL model.}
\label{fig:crpropa_ssc}
\end{figure}

The second scenario treats the secondary $\gamma$-rays as an additional component to the best-fit SSC model from Section~\ref{sectionssc} for describing the VHE emission. The normalization of the secondary $\gamma$-ray spectrum is far less constrained under this scenario, as the parameters of the UHECR spectrum are not predetermined. The choices for $\alpha_{p}$, E$_{max}$, and $\Gamma$ can vary the normalization of the secondaries by orders of magnitude while keeping the requirement on the power of the UHECRs within an acceptable range (less than the Eddington luminosity). Hence, in this scenario, we focus on the shape of the secondary $\gamma$-ray spectrum, which for a given redshift is primarily affected by the choice of the EBL model \citep{essey11}. For a selection of three EBL models that span the range of secondary spectrum shapes, the secondary $\gamma$-ray spectra are calculated and fit to the VERITAS spectral points with the requirement that the secondaries do not overshoot the VERITAS spectral points. The result for the 2013 campaign shown in Fig.~\ref{fig:crpropa_ssc} implies that the secondary $\gamma$-rays at best can only be responsible for the two highest energy VERITAS spectral points.

  \section{Conclusions}

\pks\ is a luminous high-frequency blazar, with a peak luminosity similar to the one from 3C 279, but with a peak
frequency similar to Markarian 421 (in its low state); it is currently the most distant VHE blazar detected in a non-flaring state,
and thus a unique source of VHE photons.  This work presents a systematic modeling attempt of the emission
from this source, comparing the results from different blazar emission scenarios in order to investigate the emission mechanism(s) at work in this peculiar blazar.\\

We first investigated Synchrotron Self Compton (SSC) models including both the standard one-zone and multi-zone SSC models as well as
a single-zone SSC model with contributions from external photons from the dusty torus in an External Inverse-Compton (EIC) model.
We find that the standard one-zone SSC model (which usually satisfactorily describes the SED of HBLs) cannot describe the SED of PKS 1424+240,
unless we adopt values of $\delta > 250$ and we consider a systematic bias in the $\gamma$-ray spectrum observed with Cherenkov telescopes.
However, a multi-zone SSC model alleviates the issues of the one-zone SSC scenario, and can describe the SED assuming a more
reasonable value of $\delta=30$. The EIC scenario, in which infrared photons from the dust torus are upscattered by leptons in the
relativistic jet, is also a viable alternative, requiring similar values of $\delta$.\\

We next investigated hadronic emission scenarios, including both $\gamma$-ray emission from hadrons at the source as well as secondary emission
from UHECR line-of-sight interactions. For the source emission models, no reasonable solution was found for a pure proton-synchrotron model;
however good solutions were found for proton-synchrotron emission at lower energies with synchrotron emission from secondary leptons producing
the VHE emission. Thus we show a hadronic scenario can provide a good description of the data, and naturally predicts a hardening at TeV energies. In
this case, the total power of the emitting region remains of the order of the Eddington luminosity and cannot be disfavored on this basis as is regularly done for
bright FSRQs. \\

For $\gamma$-ray emission as secondaries from UHECR line-of-sight interactions, we produced a self-consistent hadronic model as well as studied the contribution from UHECR for the one-zone SSC scenario. We find that the $\gamma$-ray emission from UHECR in the line of sight can
be dominant in the TeV regime for this source, potentially accounting for hardening at these energies.\\

To further investigate the origin of $\gamma$-ray emission from PKS 1424+240, it is fundamental to extend the spectrum in the TeV regime.
This will allow us to discriminate between models favoring hadronic emission (both in the source and in the line-of-sight) and leptonic emission.
Variability, and thus the detection of any flare, will also be a useful tool to access blazar physics. In this sense, PKS 1424+240 should be a high-priority
target for the future Cherenkov Telescope Array (CTA, Acharya et al. 2013, Sol et al. 2013). We also encourage regular monitoring by H.E.S.S., MAGIC
and VERITAS, and any $\gamma$-ray activity should trigger heavy VHE campaigns to try to extend the VHE spectrum to the highest energies. \\

 \begin{acknowledgements}
 This work has been made possible thanks to the computing centers of the Harvard-Smithsonian Center of Astrophysics, Cambridge, USA, and the Laboratoire de Physique Nucléaire et Hautes Energies, Paris, France. The informatics and support teams of both laboratories are  greatly acknowledged. Part of this work is based on archival data, software or online services provided by the ASI SCIENCE DATA CENTER (ASDC). The authors wish to thank the anonymous referee for his/her comments and suggestions which improved the quality of the paper. LFF and KS acknowledge partial support from the National Science Foundation award PHY-1407326.
 \end{acknowledgements}

 \bibliographystyle{aa}
  \bibliography{PKS1424_biblio}

   \end{document}